\documentclass[showpacs,preprint,nofootinbib,superscriptaddress,citeautoscript,eqsecnum,12pt]{article}\usepackage{latexsym}
\usepackage{amsmath}
\usepackage{amsfonts}
\usepackage{amssymb}
\usepackage{amscd}
\usepackage{bbm}
\usepackage{fancybox}
\usepackage{cite}
\usepackage{amsmath,amsfonts,amsbsy}
\usepackage{pstricks,pst-node}
\usepackage[small,bf,hang]{caption2}
\pdfoutput=1
\usepackage{graphicx}
\usepackage{epsfig}
\usepackage{psfrag}
\usepackage{comment}
\usepackage{epstopdf}
\usepackage{hyperref}

\usepackage[utf8x]{inputenc}
\usepackage{relsize}
\usepackage{dcolumn}

\usepackage{float}

\psset{unit=1.3cm,linewidth=.5pt,radius=.2}  

\usepackage{multirow}                     
\usepackage{float}                          
\usepackage{lscape}                         
\usepackage{subfigure}          
\usepackage{bm}

\newcommand{\cC}{{\cal C}}
\newcommand{\cD}{{\cal D}}

\newcommand{\cL}{{\cal L}}
\newcommand{\cM}{{\cal M}}

\newcommand{\cO}{{\cal O}}
\newcommand{\cP}{{\cal P}}

\newcommand{\beq}{\begin{equation}}
\newcommand{\eeq}{\end{equation}}
\newcommand{\bi}{\begin{itemize}}
\newcommand{\ei}{\end{itemize}}
\newcommand{\bt}{\begin{tabular}}
\newcommand{\et}{\end{tabular}}
\newcommand{\bc}{\begin{center}}
\newcommand{\ec}{\end{center}}

\def\one{{\hbox{ 1\kern-.8mm l}}}
\newcommand{\Dslash}{\not{\hbox{\kern-4pt $D$}}}
\newcommand{\pdslash}{\not{\hbox{\kern-2pt $\partial$}}}
\newcommand{\nn}{\nonumber}

\newcommand{\be}{\begin{equation}}
\newcommand{\ee}{\end{equation}}
\newcommand{\bea}{\begin{eqnarray}}
\newcommand{\eea}{\end{eqnarray}}
\newcommand{\ba}{\begin{array}}
\newcommand{\ea}{\end{array}}

\def\bbox{{\,\lower0.9pt\vbox{\hrule \hbox{\vrule height 0.2 cm
\hskip 0.2 cm \vrule height 0.2 cm}\hrule}\,}}
\newcommand{\dsl}{\pa \kern-0.5em /}

\begin{document}

\begin{titlepage}
\begin{center}

\hfill UG-14-18

\vskip 1.5cm

{\Large \bf Interacting spin-2 fields in three dimensions}

\vskip 1cm

{\bf Hamid R.~Afshar$^1$, Eric A.~Bergshoeff$^1$ and Wout Merbis$^2$}

\vskip 25pt

{\em $^1$ \hskip -.1truecm  Centre for Theoretical Physics,
University of Groningen, \\ Nijenborgh 4, 9747 AG Groningen, The
Netherlands \vskip 5pt }

{email: {\tt h.r.afshar@rug.nl, E.A.Bergshoeff@rug.nl}} \\
\vskip 10pt

{\em $^2$ \hskip -.1truecm  Institute for Theoretical Physics, Vienna University of Technology, \\
Wiedner Hauptstrasse 8-10/136, A-1040 Vienna, Austria\vskip 5pt }
{email: {\tt merbis@hep.itp.tuwien.ac.at}} \\
\vskip 10pt

\end{center}

\vskip 1.5cm

\begin{center} {\bf ABSTRACT}\\[3ex]
\end{center}

\noindent
Using the frame formulation of multi-gravity in three dimensions, we show that  demanding the presence of secondary constraints which remove the Boulware-Deser ghosts restricts the possible interaction terms of the theory and identifies invertible frame field combinations whose effective metric  may consistently couple to matter. The resulting ghost-free theories can be represented by theory graphs which are trees. In the case of three frame fields, we explicitly show that the requirement of positive masses and energies for the bulk spin-2 modes in AdS$_3$ is consistent with a positive central charge for the putative dual CFT$_2$.

\end{titlepage}

\newpage

\tableofcontents

\section{Introduction}

Theories of interacting spin-2 fields have received a lot of attention recently. While no-go theorems forbid the presence of massless interacting spin-2 fields \cite{Boulanger:2000rq,Boulanger:2000bp}, this is not the case if the interacting theories describe massive spin-2 modes. The construction of these theories should ensure the absence of unphysical degrees of freedom, like the Boulware-Deser ghost \cite{Boulware:1973my}.
Consistent ghost-free theories for a massive spin-2 field have been constructed recently \cite{deRham:2010ik,deRham:2010kj}. The construction allows for a generalization to bimetric theories of gravity \cite{Hassan:2011zd,Hassan:2011ea}, where two dynamical metrics interact through non-derivative interaction terms. For recent reviews on massive and bimetric gravity see e.g. \cite{Hinterbichler:2011tt,deRham:2014zqa}. 

A frame formulation for multiple interacting spin-2 fields was proposed in \cite{Hinterbichler:2012cn} (for more work on multi-gravity theories, see \cite{Khosravi:2011zi,Hassan:2012wt,Nomura:2012xr,Hassan:2012wr,Tamanini:2013xia,Noller:2013yja}). Instead of multiple interacting metrics, the theory is described in terms of multiple interacting frame fields. A relation with the metric formulation can be obtained under certain assumptions (see \cite{Deffayet:2012zc,Deffayet:2012nr}) and the theory was argued to be free of scalar ghost-like excitations. For two interacting frame fields in three dimensions this analysis was revisited in \cite{Bergshoeff:2013xma,Banados:2013fda,Bergshoeff:2014bia}. It was shown that demanding the absence of (possibly ghost-like) scalar excitations required an additional assumption. A linear combination of the two frame fields has to be invertible in order to derive the secondary constraints needed to remove the scalar mode.

In this paper we will investigate the absence of scalar modes in three dimensional theories with multiple ($\geq 3$) interacting frame fields. We find that only a subset of the theories considered in \cite{Hinterbichler:2012cn} possesses the necessary constraints needed to remove the additional scalar modes. In particular, we find that interaction terms mixing more than two dreibeine are forbidden. The models can then be represented by a theory graph in which vertices correspond to the different frame fields, and edges represent the interaction terms between the frame fields. We find that the ghost-free theory with $N$ frame fields must have exactly $N-1$ edges, so its theory graph is a tree. 

Once we succeed to construct a ghost free model a natural question is how to couple matter to this theory. The matter coupling should be such that it does not reintroduce the Boulware-Deser ghosts. This problem has been considered recently in \cite{Yamashita:2014fga,deRham:2014naa,Noller:2014sta,Hassan:2014gta,Schmidt-May:2014xla,deRham:2014fha,Heisenberg:2014rka,deRham:2014tga}. We show that the presence of the necessary secondary constraints in these theories is related to the presence of invertible linear combinations of frame fields. From these combinations one can form an effective dreibein, whose metric can be used to consistently couple the theory to matter, up to a cut-off scale larger than the strong coupling scale \cite{deRham:2014fha}. 

In addition to the ghost analysis, we investigate the linearized spectrum of the ghost-free multi-frame field theories around anti-de Sitter(AdS) spacetime and comment on the central charge of the putative dual conformal field theory (CFT). We show explicitly that for three frame fields (or: drei-dreibein gravity), the linear theory is a combination of two massive and one massless Fierz-Pauli Lagrangian and we compute the masses and the AdS central charge. In an earlier work \cite{Afshar:2014ffa}, we considered higher-derivative models of gravity in three dimensions and showed that obtaining positive energies and masses for the massive spin-2 fields is inconsistent with a positive dual central charge. In this work we will show that the drei-dreibein gravity model can overcome this problem and hence can consistently describe a ghost-free theory of two massive spin-2 modes in the presence of gravity, with a possibly unitary dual CFT.

This paper is organized as follows. Section \ref{MultiDG} deals with the constraint analysis of multi-metric gravity in three dimensions; after introducing its frame formalism, we review the Hamiltonian analysis of zwei-dreibein gravity, which is the frame formulation of 3D bimetric gravity. We then generalize these arguments to three frame fields (drei-dreibein gravity) in subsection \ref{sec:DDG} and to theories containing more than three frame fields (viel-dreibein gravity) in subsection \ref{sec:VDG}. Section \ref{LinDDG} is devoted to the linearized analysis and the computation of the masses and the central charge of drei-dreibein gravity around AdS. 
Finally, we summarize and conclude in section \ref{sec:Conclusion}. The explicit computation of the Poisson brackets of all the Hamiltonian constraints in drei-dreibein gravity is done in the appendix \ref{app:Hamil}. In appendix \ref{sec:ENMGlimit} we find a scaling limit of the drei-dreibein model to a sixth-order higher-derivative theory which also contains two massive modes and we obtain its  central charge via this limit.

{\bf Conventions.} We work in three spacetime dimensions and label the different frame field one-forms (dreibeine) $e_I{}^{a} = e_{I\, \mu}{}^a dx^{\mu}$ and the dualised spin connection one-forms $\omega_I{}^a= \frac12 \epsilon^{abc} \omega_{I\,bc} = \omega_{I\,\mu}{}^a dx^\mu$ by an index $I,J,\cdots$ ranging from 1 to $N$. We use Greek letters $\mu, \nu, \cdots$ for spacetime indices and $i, j, \cdots$ for the spatial indices. Lorentz indices are denoted by Latin letters $a,b, \cdots$, but these are mostly implicit and contractions of the Lorentz indices with $\eta_{ab}$ and $\epsilon_{abc}$ are denoted by dot and cross products respectively. Wedge products of the form fields are implicit throughout this work, hence $\epsilon_{abc} e^a \wedge e^b \wedge e^c$ will be denoted as $e \cdot e \times e$.

\section{Multi-gravity in three dimensions}\label{MultiDG}
The frame formulation of multi-metric gravity  was introduced in \cite{Hinterbichler:2012cn}. In this paper we restrict our attention to the three dimensional case. The multi-frame field theories of gravity can be defined in terms of a set of $N$ frame fields (dreibeine) $e_I{}^a$ and a set of $N$ dualised spin connections $\omega_I{}^a$ by the following  Lagrangian three form:
\begin{equation}\label{multi3D}
\begin{split}
L =  & - M_P \sum_{I=1}^{N} \left(  \sigma_I \, e_{I} \cdot R_I + \frac{m^2}{6}  \alpha_{I}\, e_{I} \cdot e_{I} \times e_{I} \right)  \\ 
& + \frac{m^2}{2} M_P \left\{ \sum_{ I\neq J}^{N} \beta_{IJ}\, e_{I} \cdot e_{I} \times e_{J} + \sum_{I <J < K}^{N}  \beta_{IJK}\, e_{I} \cdot e_{J} \times e_{K} \right\} \,.
\end{split}
\end{equation}
Here the dimensionless parameters $\sigma_I$ are $N$ ratios of Planck masses, from which we may always set one to unity without loss of generality. The theory further contains $N$ dimensionless cosmological parameters $\alpha_I$, $N(N-1)$ coupling constants $\beta_{IJ}$, coupling two different dreibeine and $\frac{1}{6}N(N-1)(N-2)$ coupling constants $\beta_{IJK}$ coupling three different dreibeine.\footnote{The order of indices in $\beta_{IJK}$ is inessential, while in $\beta_{IJ}$ the order is important, i.e. $\beta_{IJ} \neq \beta_{JI}$.} The mass parameter $m$ is redundant, but convenient and $M_P = \frac{1}{8\pi G}$ is the Planck mass.
The curvature $R_I$ and the torsion $T_I$ two-forms are defined as:
\begin{align}
R_I & \equiv \cD_I \omega_{I}  = d\omega_I + \frac12 \omega_I \times \omega_I\,, \\
T_I & \equiv \cD_I e_{I}  = de_I + \omega_{I} \times e_{I}\,.
\end{align}
Here $\cD_I$ denotes the covariant derivative with respect to $\omega_I$. The first line in \eqref{multi3D} is a collection of $N$ Einstein-Cartan Lagrangians, each being independently invariant under diffeomorphisms and local Lorentz rotations with the gauge field $\omega_I$. 

The presence of the interaction terms in the second line breaks these gauge symmetries to a diagonal subgroup. Hence overall, there is only one copy of diffeomorphism and local Lorentz invariance. The gauge field corresponding to the latter is the linear combination $\sum \omega_I{}^a$. In addition, the Lagrangian \eqref{multi3D} has a global symmetry; it is invariant under the action of the discrete group ($S_N$), which are the $N!$ permutations of the dreibein labels $I,J, \cdots$.

The three dimensional model \eqref{multi3D} fits into a general class of theories denoted as Chern-Simons-like theories of gravity \cite{Hohm:2012vh}. In \cite{Bergshoeff:2014bia}, the Hamiltonian analysis of this general class of theories was performed and the results also apply to this model. We will now briefly discuss some general arguments pertaining to the counting of the dynamical phase-space before moving on to a more detailed treatment of the Hamiltonian constraints in the next subsections.

This theory should describe $N-1$ massive spin-2 modes and one massless mode. A massive spin-2 mode in three dimensions has two physical degrees of freedom, and hence the desired number of degrees of freedom is $2(N-1)$, or a physical phase space of dimension $4(N-1)$. 
After a space-time decomposition of the fields, the dynamical phase space consists of the $12N$ components of the spatial parts of the dreibeine and the spin connections ($e_{I\,i}{}^a, \omega_{I\,i}{}^a$). The time components of the fields ($e_{I\,t}{}^a,\omega_{I\,t}{}^a$) act as Lagrange multipliers for $6N$ primary constraints, out of which 6 (linear combinations) are first class, corresponding to the diagonal gauge symmetries of the theory. 
In the absence of any other constraints, a counting of the physical phase-space dimensionality gives:
\begin{equation}\label{VDGcount}
12N  - 6N - 6 = 6(N-1)\,.
\end{equation}
To arrive at $4(N-1)$, we need to derive $2(N-1)$ additional second class constraints, otherwise the theory would contain additional degrees of freedom. According to Dirac's procedure for constraint Hamiltonian systems, additional constraints can follow from demanding that the primary constraints are conserved under time evolution. This leads to a set of consistency conditions which in the case of Chern-Simons-like theories can equivalently be derived on-shell by using the Bianchi and Cartan identities satisfied by the curvature and torsion two-forms (see \cite{Bergshoeff:2014bia} for more details),
\begin{align}
\cD_I R_I &=0\,,\label{bian}\\
\cD_I T_{I} &= R_I \times  e_{I}\label{cartan}\,.
\end{align}
These identities are three-form equations and hence necessarily mix the dynamical (spatial) components of the fields and the Lagrange multipliers (or, the time components)  \cite{Banados:2013fda,Bergshoeff:2014bia}. This implies that they can in principle be satisfied by restricting the Lagrange multipliers and by this logic there can never be secondary constraints. 
The situation is more subtle when one assumes some (combination) of the dreibeine to be invertible.\footnote{The assumption that (a combination of) the dreibeine is invertible is very natural in light of the interpretation of the theory as a gravitational theory. In fact, the frame formalism of three dimensional gravity is only equivalent to general relativity when one assumes the dreibein to be invertible.}   In that case, restricting the Lagrange multipliers may imply setting to zero the time components of an invertible field, leading to a contradiction.  This contradiction is avoided when the theory possesses additional constraints on the dynamical variables. These are exactly the constraints we are looking for. 

After obtaining the secondary constraints through this procedure, one should check for the presence of tertiary constraints by evaluating their Poisson brackets with the total Hamiltonian. In this case, however, no tertiary constraints can arise, as the consistency of the secondary constraints can always be satisfied by restricting Lagrange multipliers. The (assumed) invertibility of the fields is now guaranteed by the secondary constraints.

One may argue that instead of deriving the constraints followings Dirac's procedure, one could impose them by hand as part of the definition of the theory. It is obvious that we can not add these constraints with Lagrange multipliers to the action since they would change the field equations. A natural set of candidate constraints to impose are the ``symmetry conditions'' $e_I \cdot e_J = 0$, since they play a crucial role in the relation to the metric formulation --- see \cite{Hinterbichler:2012cn,Hassan:2012wt,Deffayet:2012nr,Deffayet:2012zc} for the relation between the metric and the vielbein formulation of multi-gravity theories. However, the counting argument fails for this set; there are $\frac12 N(N-1)$ symmetry conditions and $2(N-1)$ constraints needed. Furthermore, as will become clear in the following sections, after imposing these symmetry conditions, the Cartan identities \eqref{cartan} will be satisfied and a same number of other constraints follow from the Bianchi identities \eqref{bian}. 
Obviously, for $N > 2$ we have $N(N-1) > 2(N-1)$, so the theory with hand-imposed symmetry constraints is too restrictive  and we end up in the situation where the system is over-constraint. 
For generic (non-zero) values of coupling constants in \eqref{multi3D}, imposing only a subset of $2(N-1)$ out of these constraints leads to inconsistencies, as then not all equations \eqref{bian} and \eqref{cartan} are satisfied identically and one needs to restrict additional Lagrange multipliers to fulfill the consistency conditions. One could then run into the problem that the Lagrange multipliers become over-determined; not all first class constraints are imposed by a free Lagrange multiplier.

In the following we discuss this procedure for the cases $N=2$ and $N=3$  and later we generalize our findings to arbitrary $N$. We will see that for $N \geq3$, demanding the presence of additional second class constraints not only requires us to assume some (linear combination) of the dreibeine to be invertible, but also restricts the possible interaction terms of the theory.

\subsection{Zwei-dreibein gravity ($N=2$)}
The Hamiltonian form of the frame formulation of three dimensional bimetric gravity\footnote{For classical solutions in the 3D bigravity see \cite{Banados:2009it,Afshar:2009rg,Bergshoeff:2014eca,Goya:2014eya}.}, denoted as zwei-dreibein gravity (ZDG) \cite{Bergshoeff:2013xma}, was discussed at length in \cite{Banados:2013fda,Bergshoeff:2014bia}. 
Here we will review some of the subtleties which will be illustrative for the multi-dreibein generalizations considered later on. ZDG can be defined by the Lagrangian three-form:
\begin{align}
L_{\text{\tiny{ZDG}}} = & \;- M_P \left\{ \sigma e_1 \cdot R _1 + e_2 \cdot R_2 + \frac{m^2}{6}\left( \alpha_1 \, e_1 \cdot e_1 \times e_1 +  \alpha_2 \,  e_2 \cdot e_2 \times e_2 \right)\right. \nonumber \\
& \qquad\qquad\left. - \frac{m^2}{2}\left( \beta_1\, e_1 \cdot e_1 \times e_2 +\beta_2\, e_1 \cdot e_2 \times e_2 \right)\right\}\,. \label{LZDG}
\end{align}
Here the parameters $\beta_{12}$ and $\beta_{21}$ appearing in \eqref{multi3D} are denoted as $\beta_{1}$ and $\beta_{2}$ respectively, $\sigma_1 = \sigma$ and $\sigma_2$ is set to one. The counting argument in \eqref{VDGcount} reveals that the theory defined by \eqref{LZDG} generically has 3 physical degrees of freedom, two for a massive spin-2 mode in three dimensions and one for a possibly ghost-like scalar mode. In order to remove this mode we need to derive 2 secondary constraints \cite{Banados:2013fda,Bergshoeff:2014bia}.

By acting on the field equations derived from \eqref{LZDG} with an exterior derivative, one can derive a set of integrability conditions \cite{Banados:2013fda,Bergshoeff:2014bia}. 
\begin{subequations}\label{intcon}
\begin{align}
(\beta_1 e_1 + \beta_2 e_2) e_1 \cdot e_2 & = 0\,, \label{intcon1} \\
e_2 \, \omega_{12} \cdot (\beta_1 e_1 + \beta_2 e_2) - \beta_1 \omega_{12}\, e_1 \cdot e_2 & = 0\,, \label{intcon2} \\
e_1 \, \omega_{12} \cdot (\beta_1 e_1 + \beta_2 e_2) + \beta_2 \omega_{12}\, e_1 \cdot e_2 & = 0 \,. \label{intcon3}
\end{align}
\end{subequations}
where $\omega_{12} = \omega_1 - \omega_2$ and the uncontracted fields carry an implicit free Lorentz index. These equations can equivalently be derived as following from the identities \eqref{bian}-\eqref{cartan} and using the equations of motion. In the Hamiltonian formulation of the theory, the equations \eqref{intcon} are equivalent to the consistency conditions for the primary Hamiltonian constraints \cite{Bergshoeff:2014bia}. Since they are three-form equations, they will always mix Lagrange multipliers (the time components of the fields) with dynamical variables (spatial components). However, if the linear combination
\bea
\beta_1 e_1 + \beta_2 e_2\,,
\eea
has an inverse, then from \eqref{intcon} it follows that\footnote{The constraint $e_1 \cdot e_2 = 0$, sometimes called the symmetry condition, is of great importance in connecting the frame formulation to the metric form of \cite{Hassan:2011zd}. See \cite{Hinterbichler:2012cn,Hassan:2012wt,Deffayet:2012nr,Deffayet:2012zc} for further discussion on this issue.}
\begin{equation}\label{seccon}
e_1 \cdot e_2 = 0\,, \qquad \omega_{12} \cdot (\beta_1 e_1 + \beta_2 e_2) = 0\,.
\end{equation}
The spatial projection of these two equations solely involves dynamical variables and hence they constitute a set of two secondary constraints. These secondary constraints are also second class \cite{Bergshoeff:2014bia} and add to the six primary second class constraints. In total there are now 14 constraints, out of which 6 are first class and 8 second class. The counting of the dimension of the physical phase-space gives
\begin{equation}
24 - 6 \times 2 - 8 =4\,,
\end{equation}
consistent with the two helicity states of a massive spin-2 particle, which constitutes the spectrum of the linear theory. 

The above analysis shows that the presence of secondary constraints, which are needed to remove the additional scalar degree of freedom, is intrinsically related to the presence of invertible fields in the theory. This is important for the interpretation of ZDG as a theory of gravity, since the relation between a metric and a frame formulation only holds if the dreibein is invertible. In the case of ZDG we learn that in order to define a ghost-free theory of massive gravity, we should take a linear combination of the two dreibeine to be invertible. From this combination we could then construct an effective metric, defined as
\begin{equation}\label{geff}
g^{\rm eff}_{\mu\nu} = e^{\rm eff}_\mu \cdot e^{\rm eff}_\nu\qquad\text{where}\qquad e^{\rm eff}=\beta_1 e_1 + \beta_2 e_2 \,.
\end{equation}
This effective metric is invertible by assumption and it has another desirable feature. If the theory were to be coupled to matter through this effective metric, then matter loops in a quantum theory would contribute to the gravitation potential as $\sqrt{-\det(g^{\rm eff})} = \det(e^{\rm eff})$. This contribution does not reintroduce the Boulware-Deser ghost in the decoupling limit, since it only includes interaction terms which are already present in \eqref{LZDG} \cite{deRham:2014naa,Noller:2014sta,Hassan:2014gta}. However, in general the ghost is expected to re-emerge at a scale between the strong coupling scale and the Planck scale \cite{deRham:2014fha}. 

As a last note, before moving to the analysis of three or more frame fields, we introduce the theory diagrams for the ghost-free ZDG model. The theory can be represented by a diagram consisting of two nodes, representing the two dreibeine, and an edge between the nodes, representing the interaction terms $\beta_1$ and $\beta_2$. We can associate the edge connecting the two nodes with a linear combination of two dreibeine which will have an inverse,
\begin{equation}\label{tree11}
\includegraphics[height=11pt]{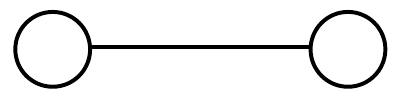} \quad \equiv \quad \beta_{1} \textrm{ and } \beta_{2}-\text{terms,\, with}\;\; \beta_1 e_1 + \beta_2 e_2 \;\;  \text{invertible.}
\end{equation}
If one restricts the coupling constants of the theory as $\beta_1 \beta_2 =0$, but one of them non-zero, then the invertibility of a single dreibein and not of any linear combination is sufficient to define a ghost-free theory. 
We can depict these interaction terms as arrows between nodes,
\begin{equation}\label{tree2}
\includegraphics[height=20pt]{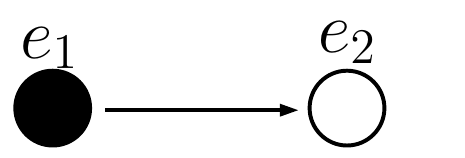}\quad\equiv \quad\beta_1-\text{term,\, with}\;\;  e_1 \;\;\text{invertible,}
\end{equation}
where the filled circle represents the invertible driebein. Note that here we picked $\beta_1$ to be non-zero, but this choice is arbitrary. An equivalent diagram can be obtained after interchanging the dreibein labels ($1\leftrightarrow2$) in \eqref{tree2}. This represents the same theory, since both diagrams are related by a discrete symmetry ($\mathbb{Z}_2$). Therefore, in the generalization of these diagrams to higher $N$ in next sections, we will ignore labeling the nodes.

\subsection{Drei-dreibein gravity ($N=3$)}\label{sec:DDG}
In this section we consider the $N=3$ extension of 3D multi-frame field gravity and introduce the drei-dreibein gravity (DDG) model which is a theory with three interacting dreibeine and three `spin connections'.\footnote{Since there is only one overall local Lorentz invariance, there is also only one `real' spin connection as the gauge field for the diagonal gauge symmetry.} 
The most general Lagrangian three-form is given by \eqref{multi3D} with $N=3$.
\begin{equation}\label{DDG}
L_{\text{\tiny{DDG}}} = - M_P \sum_{I=1}^{3}  \left( \sigma_I \, e_{I} \cdot R_I + \frac{m^2}{6} \alpha_I\, e_I{} \cdot e_I{} \times e_I{} \right) + L_{\rm int}\,,
\end{equation}
where
\begin{equation}
\begin{split}
L_{\rm int} = \,\,  & \frac{m^2}{2} M_P \Big\{ \beta_{12}\, e_1 \cdot e_1 \times e_2  +  \beta_{21}\, e_2 \cdot e_2 \times e_1  +  \beta_{13}\, e_1 \cdot e_1 \times e_3 \\ & + \beta_{31}\, e_3 \cdot e_3 \times e_1  + \beta_{23}\, e_2 \cdot e_2 \times e_3 + \beta_{32}\, e_3 \cdot e_3 \times e_2 +  \beta_{123}\, e_1 \cdot e_2 \times e_3 \Big\}\,.
\end{split}
\end{equation}
The equations of motion are
\begin{align}\label{ddgeom}
\sigma_1 R_1 =  \tfrac12 m^2 \Big[ & - \alpha_1 e_{1} \times e_{1} + \beta_{123} e_{2} \times e_{3} + \left( 2 \beta_{12} e_{1} \times e_{2} + \beta_{21} e_{2} \times e_{2} \right) \nonumber \\
&  + \left( 2 \beta_{13} e_{1} \times e_{3} + \beta_{31} e_{3} \times e_{3} \right) \Big]\,,  \nn\\
\sigma_2 R_2=  \tfrac12 m^2 \Big[ & - \alpha_2 e_{2} \times e_{2} +  \beta_{123} e_{1} \times e_{3} + \left( \beta_{12} e_{1} \times e_{1} + 2\beta_{21} e_{1} \times e_{2} \right) \\
&  + \left( 2 \beta_{23} e_{2} \times e_{3} + \beta_{32} e_{3} \times e_{3} \right) \Big]\,, \nonumber \\
\sigma_3 R_3 =  \tfrac12 m^2 \Big[ & - \alpha_3 e_{3} \times e_{3} +  \beta_{123} e_{1} \times e_{2} +  \left( \beta_{13} e_{1} \times e_{1} + 2\beta_{31} e_{1} \times e_{3} \right) \nonumber \\
&  + \left( \beta_{23} e_{2} \times e_{2} + 2\beta_{32} e_{2} \times e_{3} \right) \Big]\,,   \nn
\end{align}
together with three torsion constraints, 
\begin{equation}
T_I =  de_I + \omega_{I} \times e_{I} = 0\,,\qquad I=1,2,3\,.
\end{equation}
The three curvature two-forms satisfy three Bianchi identities \eqref{bian}, and the three torsions satisfy three Cartan identities \eqref{cartan}.

\subsubsection*{Constraint analysis}
\label{sec:constraints}
Following the general counting arguments outlined in the preceding section, in the absence of additional constraints, the dimension of the physical phase space of DDG would be 12 (see equation \eqref{VDGcount}). This implies that we need 4 secondary constraints to remove the additional degrees of freedom. From the Cartan identities \eqref{cartan} and using equations of motion in \eqref{ddgeom} we can derive three 3-form equations which are satisfied on-shell. They are
\begin{subequations}\label{CartanDDG}
\begin{align}\label{Cartan1}
\big( \beta_{12} e_{1} + \beta_{21} e_{2} + \beta_{123} e_3 \big) e_1 \cdot e_2 
 + \big(\beta_{13} e_{1} + \beta_{31} e_{3} + \beta_{123} e_2 \big) e_1 \cdot e_3 = 0\,, \\
\big( \beta_{12} e_{1}  + \beta_{21} e_{2} +  \beta_{123} e_3 \big) e_1 \cdot e_2  \label{Cartan2} - \big(  \beta_{23} e_{2} + \beta_{32} e_{3}  + \beta_{123} e_1 \big) e_2 \cdot e_3 = 0\,,  \\
\big(  \beta_{13} e_{1}  + \beta_{31} e_{3}   + \beta_{123} e_2 \big) e_1 \cdot e_3 \label{Cartan3} + \big(  \beta_{23} e_{2} + \beta_{32} e_{3} + \beta_{123} e_1 \big) e_2 \cdot e_3  = 0\,. 
\end{align}
\end{subequations}
There are only secondary constraints if these equations can be turned into 2-form identities. This can be achieved by setting to zero some of the coupling constants of the theory and assuming invertibility of (some of) the dreibeine. In order to derive two secondary constraints from this system of equations, we must restrict the parameters of the theory such that exactly one of the following combinations vanish, while the other two should have an inverse.
\begin{equation}\label{combinations}
 \beta_{12} e_{1} + \beta_{21} e_{2}+\beta_{123} e_{3}\,, \qquad \beta_{13} e_{1} + \beta_{31} e_{3} +\beta_{123} e_{2}\,, \qquad 
  \beta_{23} e_{2} + \beta_{32} e_{3} +\beta_{123} e_{1}\,.
\end{equation}
Requiring one of these combinations to vanish implies that the coupling constants appearing in the combination should vanish.\footnote{In principle, one could impose a combination to be zero as an additional constraint enforced by a Lagrange multiplier $\lambda$ in the action. This would change the field equations and only the $\lambda=0$ sub-sector of the theory would be equivalent to our original action. This condition is, however, not enforced by the new field equations and hence the resulting theory would be different. See \cite{Hohm:2012vh} for related issues in higher-derivative theories of 3D massive gravity.} This is a very strong statement since 
it implies that a non-zero $\beta_{123}$ would not lead to secondary constraints, which means interaction terms with more than two dreibeine coupled to each other are not permitted. In addition, we must set to zero (at least) two of the $\beta_{IJ}$ parameters. Here we choose to take
\bea\label{zerocouplings}
  \beta_{23}=\beta_{32}=\beta_{123}=0\,.
\eea
Of course, we could have picked any other combination in \eqref{combinations} to be zero, but this would lead to the same theory, as these choices are related to each other by a transformation of the discrete symmetry group of dreibein label permutations. Only once a specific choice has been made the $S_3$ group is broken to the subgroup which leaves this combination zero. Now, the supplementary assumption is to have,
\bea
\label{invertddg}
 \beta_{12} e_{1} + \beta_{21} e_{2} \, \qquad\text{and} \qquad\beta_{13} e_{1} + \beta_{31} e_{3} \,, 
\eea
invertible. 
The three equations \eqref{CartanDDG}  then reduce to
\begin{align}\label{SecConcov}
&( \beta_{12} e_{1} + \beta_{21} e_{2}) e_1 \cdot e_2 + (\beta_{13} e_{1} + \beta_{31} e_{3}  ) e_1 \cdot e_3 = 0\,, \nn\\
&(\beta_{13} e_{1} + \beta_{31} e_{3} ) e_1 \cdot e_3 = 0\,,  \nn\\
&(\beta_{12} e_{1} + \beta_{21} e_{2} ) e_1 \cdot e_2 = 0\,. 
\end{align}
From these equations and \eqref{invertddg} one can derive the two symmetry conditions $e_1 \cdot e_2 =0$ and $e_1 \cdot e_3 = 0$. These conditions together with the invertibility of $e_I$ and the restrictions in eqn.~\eqref{zerocouplings}, guarantee a metric formulation for the DDG theory --- see \cite{Hinterbichler:2012cn,Hassan:2012wt,Deffayet:2012nr,Deffayet:2012zc} for the multi-metric version of the multi-vielbein formulation. 
Now two secondary constraints on the spatial variables of the theory follow. They are
\begin{equation}\label{seccon1}
\varepsilon^{ij} e_{1\,i} \cdot e_{2\,j} \equiv \Delta^{e_1e_2} = 0\,, \qquad 
\varepsilon^{ij} e_{1\,i} \cdot e_{3\,j} \equiv \Delta^{e_1e_3} = 0\,.
\end{equation}
The assumption in \eqref{zerocouplings} and \eqref{invertddg} (and the ultimate theory) is invariant under the stabilizer of $e_1$ which is a reflection symmetry ($2\leftrightarrow3$). 

\begin{figure}
\centering
\includegraphics[height=0.2\textwidth]{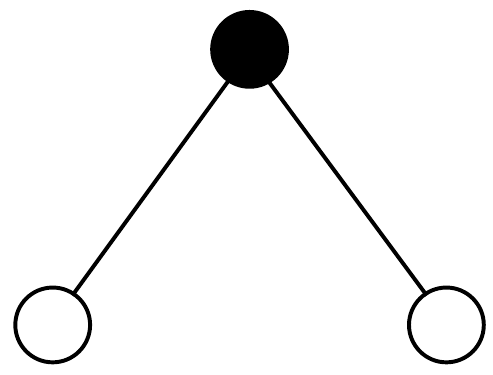}
\caption{The ghost-free interaction terms present in DDG represented in a theory diagram. The absence of interaction terms mixing more than two dreibeine demands the diagram to be a graph. The absence of ghosts further requires the graph to be a tree and the lines between the nodes represent the two invertible linear combinations of dreibeine, as in \eqref{invertddg}. In addition, the dreibein which couples to the other two dreibeine should have an inverse, denoted as a solid circle.} \label{fig:dreibein1}
\end{figure}

Let us now turn our attention to the 3 Bianchi identities \eqref{bian} and their consequence on the field equations. After acting on \eqref{ddgeom} with an exterior derivative and doing some algebra, we can derive the following 3 three-form equations:
\begin{equation}\label{ddgbianchi}
\begin{split}
& e_1 \left[\omega_{12}\cdot (\beta_{12} e_1 + \beta_{21} e_2)\right]  = 0\,, \\
& e_1 \left[\omega_{13}\cdot (\beta_{13} e_1 + \beta_{31} e_3)\right]  = 0 \,, \\
& e_2 \left[\omega_{12}\cdot (\beta_{12} e_1 + \beta_{21} e_2)\right]  + e_3 \left[\omega_{13}\cdot (\beta_{13} e_1 + \beta_{31} e_3)\right]  = 0 \,. 
\end{split}
\end{equation}
Here $\omega_{IJ} = \omega_I - \omega_J$ and we have used the parameter restrictions \eqref{zerocouplings} and the identities $e_1 \cdot e_2 = e_1 \cdot e_3 =0$. These equations are the $N=3$ generalizations of the ZDG integrability conditions \eqref{intcon2}-\eqref{intcon3}. In the ZDG case it was possible to take a linear combination of the two equations such that the effective dreibein \eqref{geff} would carry the free Lorentz index and its inverse can be used to derive another secondary constraint. In this case this is no longer possible and we see that in addition to the invertibility of the linear combinations \eqref{invertddg}, we should also require $e_1$ to have an inverse. The DDG theory can now be represented by the diagram shown in figure \ref{fig:dreibein1}. 
Another two secondary constraints can be derived from \eqref{ddgbianchi} and they read
\begin{equation}\label{seccon2}
\begin{split}
\varepsilon^{ij} \omega_{12\,i}  \cdot (\beta_{12}e_{1\,j} + \beta_{21} e_{2\,j}) \equiv \beta_{12}\Delta^{\omega_{12}e_1} + \beta_{21}\Delta^{\omega_{12} e_2} = 0\,,  \\
\varepsilon^{ij} \omega_{13\,i}  \cdot (\beta_{13}e_{1\,j} + \beta_{31} e_{3\,j}) \equiv \beta_{13}\Delta^{\omega_{13}e_1} + \beta_{31}\Delta^{\omega_{13} e_3} = 0\,. 
\end{split}
\end{equation}
These constraints together with \eqref{seccon1} are necessary and sufficient to remove all the unwanted degrees of freedom in the theory.
After adding the secondary constraints to the primary constraints, the total amount of constraints grows to 22. There are still 6 first class constraints (FCC), reflecting the six gauge symmetries present. The remaining 16 constraints are second class (SCC) and the dimension of the physical phase space, per space point, is
\begin{equation}
36\; \{\text{canonical var.}\} - 2 \times 6\; \{\text{FCC}\} - 16\;  \{\text{SCC}\} = 8\,.
\end{equation}
The total number of degrees of freedom is then four, which corresponds to the four helicity $\pm2$ modes of two massive gravitons. For more details and the explicit computation of the matrix of Poisson brackets and its rank, we refer to appendix \ref{app:Hamil}.

\begin{figure}
\centering
\subfigure[]{\includegraphics[height=0.2\textwidth]{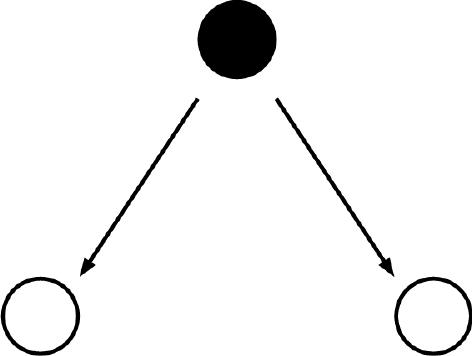}\label{fig:dreibein2a}} \quad
\subfigure[]{\includegraphics[height=0.2\textwidth]{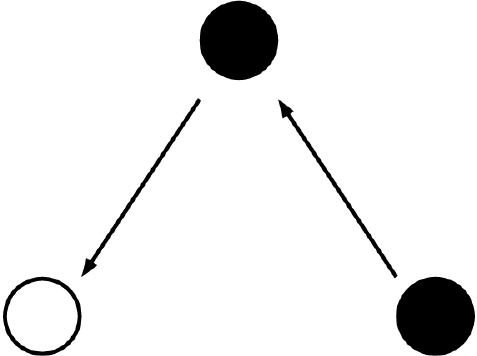}\label{fig:dreibein2b}} \quad
\subfigure[]{\includegraphics[height=0.2\textwidth]{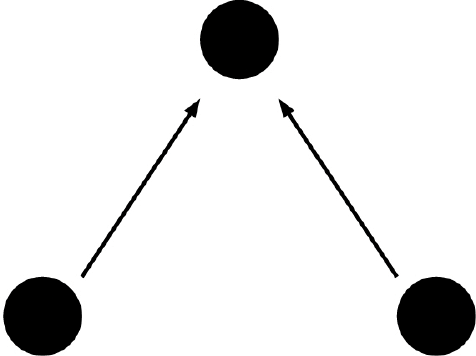}\label{fig:dreibein2c}}
\caption{Three inequivalent ghost-free interaction terms in DDG when assuming one (a), two (b) and three (c) of the original dreibeine to be invertible, while a linear combinations of them may have zero determinant. The theory graphs are oriented trees and each of them require a different number of invertible frame fields, denoted here by solid circles} \label{fig:dreibein2}
\end{figure}

There are several special cases where the invertibility of only the original dreibeine, and not of some linear combination of them, is sufficient. 
For instance, if we assume invertibility of only $e_1$, there is a unique parameter choice which leads to secondary constraints
\begin{equation}\label{paramchoice}
\beta_{12} \neq 0\, \qquad\text{and}\qquad \beta_{13} \neq 0\,,
\end{equation}
while all other $\beta$-parameters should be set to zero. 
A similar choice of parameters exists if we take $e_2$ or $e_3$ to be invertible, as is indicated in figure \ref{fig:dreibein2a}. 

The parameter choice \eqref{paramchoice} is unique if we assume the invertibility of only one dreibein. If instead there are two invertible dreibeine, then there are two other possibilities of choosing two non-zero $\beta$-parameters:
\begin{equation}\label{paramchoice2}
\beta_{12}\,, \beta_{31} \neq 0\,,\qquad
\beta_{13}\,, \beta_{21} \neq 0\,.
\end{equation}
These choices are equivalent theories, as they are related to each other by the residual reflection symmetry which transforms the dreibein labels ($2 \leftrightarrow 3$). The corresponding diagram is depicted in figure \ref{fig:dreibein2b}. Figure \ref{fig:dreibein2c} corresponds to a ghost-free theory with all three dreibeine invertible and
\begin{equation}
\beta_{31} ,\beta_{21} \neq 0\,.
\end{equation}
When assuming all three dreibeine to be invertible, the number of ghost-free DDG theories becomes equal to the number of oriented trees with three unlabeled nodes.\footnote{Note that when not all of the dreibeine are invertible, like in figure \ref{fig:dreibein2a} and \ref{fig:dreibein2b}, the arrows are always pointing from the solid circle to the empty ones. This limits the number of oriented diagrams in these cases, i.e. one for the first diagram and two for the second diagram.}

As was discussed in \cite{deRham:2014naa}, multi-gravity theories can be coupled to matter by making use of an effective metric. This effective metric should be constructed such that its determinant does not reintroduce the interaction terms that have been set to zero. Hence matter can be coupled to any of the original dreibeine which are assumed to be invertible. In addition to this, an effective metric can be constructed out of two of the three dreibeine. An effective metric containing all three dreibeine is not allowed, since its determinant would reintroduce the $\beta_{123}$ term. Coincidently, there are also two linear combinations in DDG which are required to be invertible in order to derive the necessary secondary constraints. They are related by a reflection symmetry (${\mathbb Z}_2$) transformation ($2\leftrightarrow3$) so they can be regarded as the effective dreibein of the theory up to a symmetry transformation, 
\bea\label{e3eff}
e_{\rm eff}=\beta_{1J} e_{1} + \beta_{J1} e_{J} \qquad\text{with}\qquad J=2,3\,.
\eea
The effective metric constructed from this dreibein is a good candidate for an effective field theory coupling to matter, since it is invertible by assumption and its determinant does not reintroduce the couplings which were set to zero in \eqref{zerocouplings}, hence no ghosts reappear in the decoupling limit. 
Moreover the single internal dreibein (in this case $e_1$) should also be invertible and one may couple matter directly to the metric constructed from it. This means that in contrast to the $N=2$ case, in the $N=3$  case with the most general allowed interaction terms, there are two choices of invertible metrics for matter coupling; either the effective linear combination \eqref{e3eff} or the internal dreibein itself ($e_1$ in this case). 

To summarize, the analysis of the presence of secondary constraints in DDG severely restricts the number of allowed interaction terms. In full generality, to derive the secondary constraints needed to remove the extra degrees of freedom, we must:
\begin{itemize}
\item Set to zero $\beta_{123}$, i.e. there is no interaction term mixing three dreibeine. (The theory can be pictorially represented as a graph)
\item Assume invertibility of two out of the three linear combinations in \eqref{combinations}, while the parameters of the theory should be restricted such that the third combination vanishes. (The theory graph is a tree)
\item In addition to the invertible linear combinations, the dreibein coupling to two other dreibeine should be invertible. (Internal nodes should correspond to invertible frame fields)
\end{itemize}
The second rule implies that only one dreibein is allowed to couple to the other two dreibeine, i.e. there can be no loops in the diagrams in figure \ref{fig:dreibein2}. This is compatible with the analysis of ref.~\cite{Nomura:2012xr}, obtained by different means.
In a restricted case where the original dreibeine themselves are assumed to be invertible, and not necessarily their linear combinations, the allowed theory graphs become all possible oriented trees with 3 nodes in figure \ref{fig:dreibein2}. The internal node should always be invertible.

\subsection{Viel-dreibein gravity}
\label{sec:VDG}
The constraint analysis of drei-dreibein gravity can be extended to a theory with an arbitrary number of interacting dreibeine. From the DDG analysis, we know that interaction terms mixing more than two dreibeine do not lead to secondary constraints. To investigate the secondary constraints in a viel-dreibein theory with $N$ interacting dreibeine, we only consider interaction terms involving two dreibeine and define the Lagrangian as:
\begin{equation}\label{VDG}
L_{\text{\tiny{VDG}}} = - M_P \sum_{I=1}^{N}  \left( \sigma_I \, e_{I} \cdot R_I + \frac{m^2}{6} \alpha_I\, e_I \cdot e_I \times e_I \right) + L_{\rm int}\,,
\end{equation}
where
\begin{equation}
L_{\rm int} = \,\, \frac{m^2}{2} M_P  \sum_{ I\neq J}^{N} \beta_{IJ}\, e_{I} \cdot e_{I} \times e_{J}\,.
\end{equation}
From the Bianchi and Cartan identities \eqref{bian}-\eqref{cartan}, we can derive secondary constraints if we can use the inverse of some (combination) of the dreibeine to construct two-form identities out of them. From the counting argument in \eqref{VDGcount} we know that in order for the theory to describe $N-1$ massive spin-2 modes and one massless mode,
we need to derive $2(N-1)$ secondary constraints from the Bianchi and Cartan identities. Let us first consider the Cartan identities for this model. They are $N$ equations, each containing $N-1$ terms
\begin{align} \label{VDGCartan}
 &(\beta_{12} e_1 + \beta_{21} e_2)e_1 \cdot e_2 + (\beta_{13} e_1 + \beta_{31} e_3)e_1 \cdot e_3  + \ldots 
 +  (\beta_{1N}e_{1} + \beta_{N1}e_{N})e_1 \cdot e_N  = 0\,, \nn\\[.2truecm]
 &(\beta_{12} e_1 + \beta_{21} e_2)e_2 \cdot e_1 +  (\beta_{23} e_2 + \beta_{32} e_3)e_2 \cdot e_3  + \ldots 
  +  (\beta_{2N}e_{2} + \beta_{N2}e_{N})e_2 \cdot e_N = 0\,, \nn\\
&\;\; \vdots \\  \nn
&(\beta_{1N}e_1 + \beta_{N1}e_{N}) e_N \cdot e_1   +
\ldots + (\beta_{NN-1}  e_{N}  +  \beta_{N-1N} e_{N-1}) e_N \cdot e_{N-1}  = 0\,.
\end{align}
where the uncontracted form fields have a free Lorentz index.
To derive $N-1$ secondary constraints from \eqref{VDGCartan}, we need to constrain the parameters such that $N-1$ of these involve solely an invertible combination of dreibeine carrying a free Lorentz index (the terms in the parenthesis). Then we can use the inverse of this combination of dreibeine to construct a two-form equation, whose spatial projection is a secondary constraint. In other words, out of the $\frac12 N(N-1)$ combinations
\begin{equation}\label{VDGlincom}
\beta_{IJ} e_I + \beta_{JI} e_J\, \qquad \text{with } \qquad I \neq J\,,
\end{equation}
we need that $N-1$ have an inverse, while the others all vanish. 

We now turn our attention to the $N$ Bianchi identities $\cD_I R_I = 0$. They can be written, in full generality, as
\begin{align}\label{VDGBianchi} \nonumber
& \beta_{12} \omega_{12}\, e_1 \cdot e_2 + \beta_{13} \omega_{13} \, e_1 \cdot e_3 + \ldots + \beta_{1N} \omega_{1N}\, e_1 \cdot e_N + e_2\, (\beta_{12} e_1 + \beta_{21} e_2) \cdot \omega_{12}  \\ \nonumber
&  + e_3\, (\beta_{13} e_1 + \beta_{31} e_3) \cdot \omega_{13} +  \ldots  + e_N\, ( \beta_{1N} e_1 + \beta_{N1} e_N) \cdot \omega_{1N} = 0 \,, \\[.3truecm] \nonumber
& \beta_{21} \omega_{21}\, e_2 \cdot e_1 + \beta_{23} \omega_{23}\, e_2 \cdot e_3 + \ldots + \beta_{2N} \omega_{2N}\, e_2 \cdot e_N  + e_1\, (\beta_{12} e_1 + \beta_{21} e_2) \cdot \omega_{21}  \\ \nonumber
& + e_3\, (\beta_{32} e_3 + \beta_{23} e_2) \cdot \omega_{23} + \ldots  + e_N\, ( \beta_{2N} e_2 + \beta_{N2} e_N) \cdot \omega_{2N} = 0 \,, \\
 & \qquad \qquad \qquad \qquad \qquad \qquad \qquad \qquad \qquad 
 \vdots \\ \nonumber
& \beta_{N1} \omega_{N1}\, e_N \cdot e_1 + \ldots + \beta_{NN-1} \omega_{NN-1}\, e_N \cdot e_{N-1}  \\ \nonumber
& + e_1\, (\beta_{1N} e_1 + \beta_{N1} e_N) \cdot \omega_{N1} + \ldots  + e_{N-1}\, ( \beta_{N-1N} e_{N-1} + \beta_{NN-1} e_N) \cdot \omega_{NN-1} = 0 \,.
\end{align}
For simplicity, we will restrict our attention here to a special case of the assumption \eqref{VDGlincom} where only a single dreibein is assumed to be invertible. We may pick the invertible dreibein to be $e_1$ without loss of generality. There is then a single parameter choice for which eqn.~\eqref{VDGCartan} leads to secondary constraints. This is 
\begin{equation}\label{VDGparam}
\beta_{1J} \neq 0 \, \qquad \text{for } \qquad J = 2, \ldots, N\,,
\end{equation}
and all other $\beta$-parameters vanish. The theory graph for this model is a tree where all edges are arrows pointing away from $e_1$.  After this parameter restriction, $N-1$ secondary constraints can be derived by acting with $e_1^{-1}$ on \eqref{VDGCartan} and taking the spatial part. They are
\begin{equation}
\label{VDGconstraints1}
\varepsilon^{ij} e_{1\,i} \cdot e_{J\,j} = \Delta^{e_1e_J} = 0\, \qquad \text{for } \quad J = 2, \ldots, N\,.
\end{equation}
The parameter constraint \eqref{VDGparam} and the identities $e_1 \cdot e_J = 0$ reduce the set of Bianchi identities \eqref{VDGBianchi} to:
\begin{equation}
\begin{split}
\beta_{12} e_2\, e_1 \cdot \omega_{12} + \ldots + \beta_{1N} e_N e_1 \cdot \omega_{1N} & = 0\,, \\[.2truecm]
\beta_{12}  e_1\, e_1 \cdot \omega_{21} & = 0\,, \\[.2truecm]
\beta_{13}  e_1\, e_1 \cdot \omega_{31} & = 0 \,, \\
 & \vdots \\
\beta_{1N} e_1\, e_1 \cdot \omega_{N1} & = 0 \,.
\end{split}
\end{equation} 
The assumed invertibility of $e_1$ is sufficient to derive another $N-1$ secondary constraints. They are:
\begin{equation}\label{VDGconstraints2}
\varepsilon^{ij} e_{1\,i} \cdot (\omega_{1\,j} - \omega_{J\,j}) = 0 \qquad \text{for} \quad J = 2, \ldots, N\,.
\end{equation}
Provided that the secondary constraints \eqref{VDGconstraints1} and \eqref{VDGconstraints2} are second-class, these remove an additional $2(N-1)$ components from the counting performed in \eqref{VDGcount}, leading to a physical phase space of dimension $4(N-1)$, corresponding to $2(N-1)$ degrees of freedom, the correct number of degrees of freedom to account for the two helicity states of $N-1$ massive spin-2 modes. 

In a similar fashion, secondary constraints can be derived when assuming more or all dreibeine to be invertible, or when assuming $N-1$ of the linear combinations in \eqref{VDGlincom} to be invertible. In the latter case, to derive the necessary constraints from the Bianchi identities \eqref{VDGBianchi}, we need that any dreibein which couples to more than one other dreibein should have an inverse. In other words, any internal node in the theory graph should correspond to an invertible dreibein.

\begin{figure}
\centering
\subfigure[$N=4$]{
\includegraphics[height=0.14\textwidth]{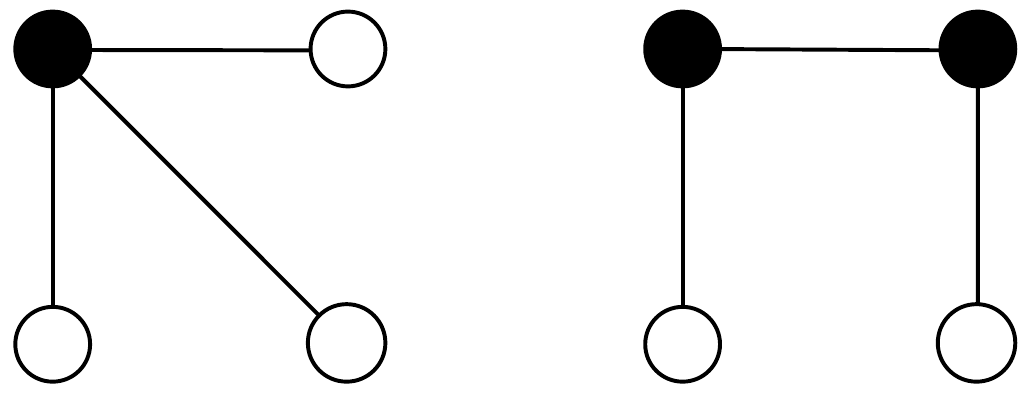}\label{fig:vdgtrees} 
} \\
\subfigure[$N=5$]{\includegraphics[height=0.14\textwidth]{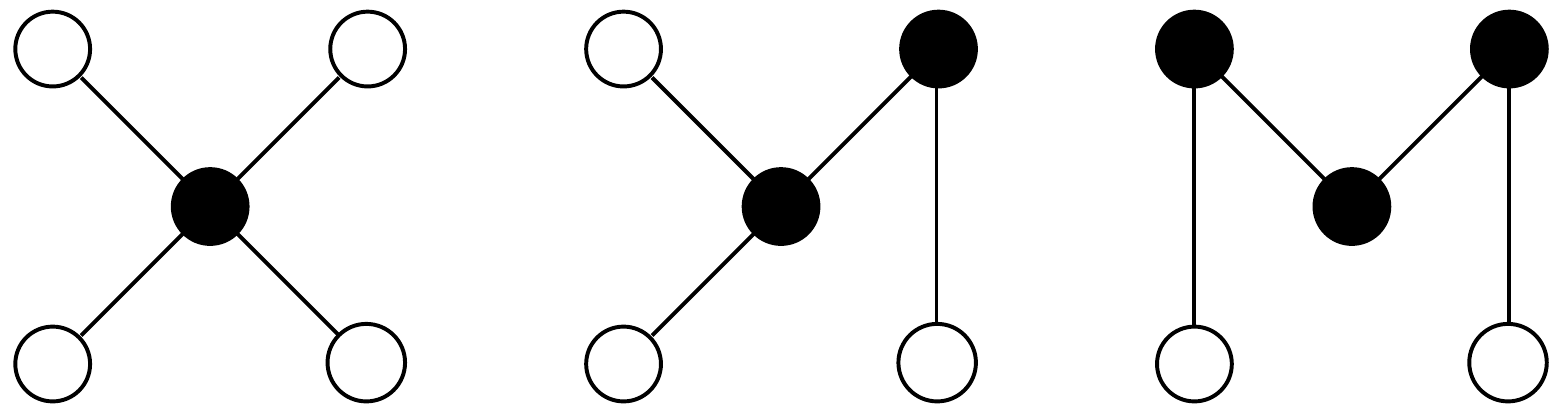}\label{fig:fdgtrees}}
\caption{The ghost-free theory diagrams for $N=4$ (a) and $N=5$ (b). In general the ghost-free VDG theory diagram is a tree where all the edges correspond to invertible linear combinations of frame fields and nodes with degree $\geq 2$ (internal nodes) correspond to invertible frame fields.} \label{fig:dreibein3}
\end{figure}

To conclude, in all of the ghost-free viel-dreibein gravity models, the parameters of the theory must be restricted such that in a theory diagram similar to figure \ref{fig:dreibein2}, but now containing $N$ nodes, the following rules hold
\begin{itemize}
\item Interaction terms with more than two dreibeine are not allowed ($\beta_{IJK}=0$ in \eqref{multi3D} and the theory diagram is a graph).
\item There must be exactly $N-1$ lines connecting the nodes and no node should be disconnected. (The theory graph is a tree)
\item Any internal dreibein (coupling to more than one other dreibein) should have an inverse (Nodes of degree $\geq 2$ must correspond to invertible frame fields).
\end{itemize}
The second point excludes the possibility to form loops in the diagrams. The mathematical tool to display these allowed diagrams is the notion of unoriented unlabeled trees; the diagrams for $N=4,5$ are shown in figure \ref{fig:dreibein3}.

Whenever the original dreibeine are assumed to be invertible, and not necessarily their linear combinations, then the interaction terms are restricted even further and the number of ghost-free VDG models equals  the number of different oriented trees with $N$ unlabeled nodes.\footnote{The number of unoriented and oriented trees with $N=1,2,3,\cdots$ unlabeled nodes, is
$1, 1, 1, 2, 3, 6, \cdots$ and $1,1,3,8,27,91,\cdots$ respectively --- see \href{http://oeis.org/A000055}{A000055} and \href{http://oeis.org/A000238}{A000238} at ``The On-Line Encyclopedia of Integer Sequences''.}.

Any multi-gravity theory constructed in this way will have the correct number of degrees of freedom to describe $N-1$ massive gravitons interacting with each other, provided that the secondary constraints derived here are also second-class and that the consistency conditions of the secondary constraints do not introduce any further tertiary constraints \cite{Bergshoeff:2014bia}. This has been checked explicitly for VDG with $N=2$ (ZDG) in \cite{Bergshoeff:2014bia} and $N=3$ (DDG) in the appendix \ref{app:Hamil}.

\section{Linearized theory}\label{LinDDG}
This section deals with the linearized spectrum of DDG around a maximally symmetric background. We show explicitly that the linearized theory describes two massive spin-2 modes and compute the Fierz-Pauli masses. We then analyze further restrictions on the ghost-free DDG parameter space by demanding the Fierz-Pauli masses, the energy of the massive modes and the central charge of the putative dual CFT to be positive.   

Before selecting a specific ghost-free theory, we analyze the linearized theory around a common maximally symmetric background with cosmological constant $\Lambda$, described by the dreibein $\bar{e}$ and the spin-connection $\bar{\omega}$. 
\begin{align}\label{fluctuations}
e_{I} & =  \gamma_I(\bar{e} + \kappa k_{I})\,, &
\omega_{I} & = \bar{\omega} + \kappa v_{I}\,.
\end{align}
Where $\gamma_I$ is the arbitrary scale parameter of $e_I$.
In the rest we put $\gamma_1=1$ by rescaling $\bar{e}$ and $k_1$.
Plugging \eqref{fluctuations} into the Lagrangian \eqref{DDG} leads to the following expansion,
\bea\label{LLL}
L_{\text{\tiny{DDG}}}=L^{(0)}_{\text{\tiny{DDG}}}+\kappa L^{(1)}_{\text{\tiny{DDG}}}+\kappa^2 L^{(2)}_{\text{\tiny{DDG}}}+\cdots\,.
\eea
The linear terms in the $\kappa$-expansion cancel when
\begin{align}\label{gammas}
\sigma_1 \frac{\Lambda}{m^2}  & =   2\beta_{12}\gamma_2 + \beta_{21}\gamma_2^2 + 2\beta_{13}\gamma_3 + \beta_{31}\gamma_3^2 - \alpha_1  \,, \nonumber \\
 \sigma_2 \frac{\Lambda}{m^2}  &=   \beta_{12} + 2\beta_{21}\gamma_2 + 2\beta_{23}\gamma_2\gamma_3 + \beta_{32}\gamma_3^2 - \gamma_2^2 \alpha_2\,, \\
 \sigma_3 \frac{\Lambda}{m^2} &=   \beta_{13} + 2\beta_{31}\gamma_3 + \beta_{23}\gamma_2^2 + 2\beta_{32}\gamma_2\gamma_3 - \alpha_3\gamma_3^2 \,. \nonumber
\end{align}
where $\beta_{123}$ is put to zero by assumption.
These three equations fix $\gamma_2, \gamma_3$ and $\Lambda$ in terms of the DDG parameters.
The quadratic Lagrangian for the fluctuations $k_I$ and $v_{I}$ is
\begin{align}\label{Lquad}
L^{(2)}_{\text{\tiny{DDG}}} = & - M_P \sum_{I=1}^{3} \sigma_I \gamma_I  \Big\{ k_{I} \cdot \bar{\cD}v_{I} + \tfrac12 \bar{e} \cdot \left( v_{I} \times v_{I} - \Lambda k_{I} \times k_I \right) \Big\} \nonumber \\
 &- \frac{m^2}{2} M_P \, \bar{e} \cdot  \Big\{ \gamma_2 (\beta_{12} + \gamma_2 \beta_{21}) (k_1 - k_2) \times (k_1 - k_2)  \\
&\qquad\qquad\qquad +  \gamma_3(\beta_{13} + \gamma_3\beta_{31})  (k_1 - k_3) \times (k_1 - k_3)  \nonumber \\ 
&\qquad\qquad\qquad  + \gamma_2 \gamma_3(\gamma_2 \beta_{23} + \gamma_3 \beta_{32})  (k_2 - k_3) \times (k_2 - k_3) \Big\} \,. \nonumber
\end{align}
In the last section we showed that one of the three combinations in \eqref{combinations} should vanish by a restriction of the parameters. We take $\beta_{13} = \beta_{31} = 0$\footnote{This implies that only one of the three dreibeine --- $e_2$ in this case --- can couple to the other ones simultaneously.}, which means, the $(k_1-k_3)^2$ term is not present in the quadratic Lagrangian. 
Up to a global symmetry relabeling the dreibein indices $I =1,2,3$, there are always two non-diagonal mass terms in the linearized theory \eqref{Lquad}.
This means that the specific ghost-free parameter choice does not influence the linearized spectrum. For simplicity, we will proceed the linearized analysis by also taking $\beta_{21} =0$ and $\beta_{32} = 0$, as they can be absorbed in $\beta_{12}$ and $\beta_{23}$.

\subsection{Mass eigenstates}
The above quadratic Lagrangian has diagonal kinetic terms for the fields $k_I$ and $v_I$, and contains mass-terms for the differences $k_1 - k_2$ and $k_2-k_3$. Here we diagonalize the theory and write it in terms of a massless field and two massive fields. 
The first step is to define two new fields equal to the difference appearing in the mass terms
\begin{align}
f_1 = k_1 - k_2\,, && f_2 = k_2 - k_3\,, \\
w_1 = v_1 - v_2 \,, && w_2 = v_2 - v_3\,.
\end{align}
We also redefine the fields $k_2$ and $v_2$ as
\begin{equation}
\begin{split}
k_2 = k_{(0)} - \frac{\sigma_1}{\gamma_{\rm crit}} f_1 + \frac{\sigma_3\gamma_3 }{\gamma_{\rm crit}} f_2 \,, \qquad
v_2 = v_{(0)} - \frac{\sigma_1 }{\gamma_{\rm crit}} w_1 + \frac{\sigma_3\gamma_3}{\gamma_{\rm crit}} w_2 \,,
\end{split}
\end{equation}
where
\begin{equation}
\label{gammacrit}
\gamma_{\rm crit} = \sigma_1 + \sigma_2 \gamma_2 + \sigma_3 \gamma_3 \,.
\end{equation} 
Assuming $ \gamma_{\rm crit} \neq 0$, the quadratic Lagrangian becomes:
\begin{align}\label{Lquad2}
L^{(2)}_{\text{\tiny{DDG}}} & =  - \gamma_{\rm crit}\, M_P \Big\{ k_{(0)} \cdot \bar{\cD}v_{(0)} + \tfrac12 \bar{e} \cdot \left( v_{(0)} \times v_{(0)} - \Lambda k_{(0)} \times k_{(0)} \right) \Big\} \nonumber \\
& - \sigma_1 (\gamma_2 \sigma_2 + \gamma_3\sigma_3)\gamma_{\rm crit}^{-1}\, M_P\Big\{ f_{1} \cdot \bar{\cD}w_{1} + \tfrac12 \bar{e} \cdot \left( w_{1} \times w_{1} - \Lambda f_{1} \times f_1 \right) \Big\} \nonumber\\
& - \gamma_3\sigma_3 (\sigma_1 + \gamma_2\sigma_2 )\gamma_{\rm crit}^{-1}\,M_P  \Big\{ f_{2} \cdot \bar{\cD}w_{2} + \tfrac12  \bar{e} \cdot \left( w_{2} \times w_{2} - \Lambda f_{2} \times f_2 \right) \Big\}   \\
& - \gamma_3 \sigma_1 \sigma_3 \gamma_{\rm crit}^{-1}\,M_P \Big\{ f_{1} \cdot \bar{\cD}w_{2} + f_{2} \cdot \bar{\cD}w_{1}  +  \bar{e} \cdot \left( w_{1} \times w_{2} - \Lambda f_{1} \times f_2 \right) \Big\} \nonumber \\
& - \frac{ m^2}{2} M_P \gamma_2 \bar{e} \cdot  \left( \beta_{12}  f_1 \times f_1  + \gamma_2 \gamma_3 \beta_{23} f_2 \times f_2 \right) \,. \nonumber
\end{align}
We have now traded an off-diagonal mass-term for off-diagonal kinetic terms, but at least we are able to identify the Lagrangian for the massless spin-2 mode, represented in the first line of \eqref{Lquad2}. 

We now diagonalize the massive part of the linearized Lagrangian, which corresponds to the last four lines of \eqref{Lquad2}. We redefine the fields $f_1, f_2,w_1$ and $w_2$ as a linear combination of two massive spin-2 modes
\begin{equation}
\begin{split}
f_1 =   k_{(+)} +   k_{(-)}\,, \qquad &
f_2 = A_+ k_{(+)} + A_- k_{(-)} \,, \\
w_1 =   v_{(+)} +   v_{(-)} \,, \qquad &
w_2 = A_+ v_{(+)} + A_- v_{(-)} \,.
\end{split}
\end{equation}
The massive part of \eqref{Lquad2} is diagonal if the (non-zero) dimensionless coefficients $A_{\pm}$ solutions of
\begin{equation}
\begin{split}
A_{+}A_{-}=\frac{\beta_{12}}{\beta_{23}\gamma_2\gamma_3}\,,\qquad A_{+}+A_{-}+ 1=
  \frac{\beta_{12}}{\beta_{23} \gamma_2\gamma_3}\left(1 + \frac{\gamma_2\sigma_2 }{\sigma_1}\right) - \frac{\gamma_2\sigma_2}{\gamma_3\sigma_3}  \,.
\end{split}
\end{equation}
The quadratic Lagrangian factorizes into a part describing a linearized massless spin-2 plus 2 massive Fierz-Pauli Lagrangians. 
\begin{align}\label{Lquad3}
L^{(2)}_{\text{\tiny{DDG}}} =  -  M_P \Big\{ \gamma_{\rm crit} \, L_{\text{\tiny{FP}}}(k_{(0)},0)+\cC_+ \, L_{\text{\tiny{FP}}}(k_{(+)},\cM_+) + \cC_- \,  L_{\text{\tiny{FP}}}(k_{(-)},\cM_-)  
   \Big\}\,, \nonumber
\end{align}
where
\begin{equation}
L_{\text{\tiny{FP}}}(k, \cM) =  k \cdot \bar{\cD}v + \frac12 \bar{e} \cdot \left( v \times v - (\Lambda - \cM^2) k \times k \right) \,,
\end{equation}
is the Fierz-Pauli Lagrangian in a first order form.

The coefficients in front of the kinetic terms of the Fierz-Pauli Lagrangian $\cC_\pm$
 and the two Fierz-Pauli masses, $\cM_\pm$, belonging to the massive modes $k_{(\pm)}$ respectively, are given by:
\begin{align}
\cM_\pm^2 &  = \frac{\left( \beta _{12} + \gamma_2 \gamma_3 \beta _{23} A_\pm^2  \right)\gamma_{\rm crit} \gamma_2 }{\cC_\pm}m^2\,, \label{DDGmass}\\
\cC_\pm &=   \gamma_2\sigma_2 ( \sigma_1 + \sigma_3\gamma_3 A_\pm^2) + \sigma_1 \sigma_3 \gamma_3   (1+A_\pm)^2 \,. \label{DDGcs}
\end{align}
Positivity of the coefficients in front of the massive kinetic terms, $\cC_\pm>0$, 
ensures that these massive modes are not ghosts. On the other hand, although the massless kinetic term does not propagate any physical local degree of freedom, positivity of its coefficient, $\gamma_{\rm crit}>0$, 
is necessary for having positive charges in the theory, such as black hole masses or the AdS central charge\footnote{For $\Lambda =  - 1/ \ell^2$, the AdS central charge in 3D multi-gravity theories has a simple form \cite{Bergshoeff:2013xma},
\begin{equation}\label{DDGcc}
c_{L/R} = 12\pi \ell M_P \gamma_{\rm crit} \,\qquad\text{where}\qquad \gamma_{\rm crit} = \sum_I^N \sigma_I\gamma_I\,.
\end{equation}}. These are the necessary conditions for having a unitary theory.
A sufficient (but not necessary) condition to fulfill these requirements is 
\bea
\gamma_I,\sigma_I>0\,.
\eea
together with $\beta_{12},\beta_{23}>0$, which also guarantees the Fierz-Pauli masses \eqref{DDGmass} to be positive.  
This shows that there is a continuous range of parameters defining a class of DDG models with good bulk and boundary properties.

\section{Conclusion and discussion}
\label{sec:Conclusion}

In this paper we discussed the three dimensional version of the multi-frame gravity theories considered in \cite{Hinterbichler:2012cn}. This theory is an extension of the first-order formulation of three dimensional bigravity, or zwei-dreibein gravity (ZDG) \cite{Bergshoeff:2013xma}.   We have revisited the counting of the dimensionality of the physical phase space and showed that demanding the absence of Boulware-Deser modes requires us to impose certain restrictions on the possible interactions of the theory. 
These restrictions have implications to the coupling to matter. Interaction terms mixing three dreibeine are not allowed for a ghost-free theory. As a consequence, the effective metric to which matter can couple can contain at most two dreibeine, otherwise the determinant of the effective dreibein would reintroduce the forbidden interaction terms. Below we elaborate and discuss these results.

{\bf Ghost freedom.} Since no interaction with more than two dreibeine is allowed, we can represent the theory pictorially by a graph where the frame fields are denoted by vertices and the interaction terms by edges between them. 
The theories free of Boulware-Deser ghosts are those connected graphs\footnote{Disconnected vertices represent isolated Einstein-Cartan theories which are completely decoupled.}  in which for $N$ vertices there are exactly $N-1$ edges --- which imply that the theory graph cannot have closed loops. 
Such a graph is a {\it tree}. In the most general case there are two interaction terms mixing two dreibeine and each vertex of the tree represents both terms --- see the diagram \eqref{tree11}. For displaying a restricted class of ghost-free models in which only one of these terms is present we can use the notion of {\it oriented tree}  --- see the diagram \eqref{tree2}.  

In addition to these restrictions on the type of interactions, we also need to assume the invertibility of some of the frame fields. In the most restricted models (oriented trees), the invertibility of a single dreibein is sufficient to derive the necessary Hamiltonian constraints. As in any theory of gravity the metric is invertible, this invertible dreibein can be thought of as the physical dreibein of the theory. 

In the most general case (unoriented trees), however, we need that $N-1$ out of $\frac{1}{2}N(N-1)$ linear combinations of two frame fields, $\beta_{IJ} e_I + \beta_{JI} e_J$, 
have an inverse, while the others should vanish. 
The $N-1$ non-zero and invertible linear combinations are related to each other by a subgroup of the permutation group $S_N$ which leaves the rest of couplings to be zero.  
For $N=3$, the two invertible combinations are related by interchanging labels by a ${\Bbb Z}_2$ symmetry (the subgroup of $S_3$ that leaves the other combination zero). This is now the symmetry left in the action after removing the unwanted interactions. 
This relabeling of dreibein indices which leaves the theory invariant can take one invertible combination into the other, so one can choose one of the two invertible combinations to correspond to a physical metric without loss of generality.

{\bf Coupling to matter.}
We can consider the invertible combinations as the effective dreibein whose metric can couple to matter,
\begin{equation}
e_{\rm eff}=\hat{\beta}_{IJ} e_I + \hat{\beta}_{JI} e_J \,\qquad\text{with}\qquad I \neq J\,,
\end{equation}
where the hat refers to the $N-1$ nonzero couplings. The reason is that, the quantity $\sqrt{-\det(g_{\rm eff})} = \det(e_{\rm eff})$  is well-defined because $e_{\rm eff}$ is invertible by assumption. Furthermore, $\det(e_{\rm eff})$ does not reintroduce interaction terms which were set to zero for ghost freedom.\footnote{For issues involving coupling to matter in bimetric and multi-metric theories, see for instance \cite{Yamashita:2014fga,deRham:2014naa,Noller:2014sta,Hassan:2014gta,Schmidt-May:2014xla,deRham:2014fha,Heisenberg:2014rka,deRham:2014tga}.} Note, however, that as shown in \cite{deRham:2014fha}, in general the ghost may re-emerge at a scale between the strong coupling scale and the Planck scale. This would imply that the aforementioned coupling to matter may only make sense as an effective field theory below some cut-of scale.

In both the oriented and the unoriented case, the `internal' vertices represent invertible dreibeine. This means that even in the most general case (unoriented) we also have the possibility of coupling to matter via metrics corresponding to the internal vertices. 
In all cases, the number of inequivalent ways of coupling matter to multi-gravity theories equals the number of different inequivalent invertible combination of dreibeine.

In the unoriented case, diagrammatically, this number corresponds to the different {\it inequivalent edges} plus the number of {\it inequivalent internal vertices} where `inequivalent' means, up to the discrete symmetries of the diagram. In the ZDG case ($N=2$) \eqref{tree11}, there is a unique invertible linear combination \eqref{geff}. In the DDG case ($N=3$) there are two inequivalent choices, as shown in figure \ref{fig:dreibein1}. There is one internal vertex and one inequivalent edge (the other edge represents the same invertible combination due to symmetry). In the $N=4$ case shown in figure \ref{fig:vdgtrees}, there are two classes of diagrams where in the left diagram there are two consistent ways of coupling to matter and in the right one there are three consistent ways.
In the oriented case, the number of inequivalent ways for matter coupling is the number of inequivalent filled circles.

In any case, when one considers multi gravity theories with $N>2$, the arguments presented in this work do not single out a unique choice for the physical metric. It would be interesting to investigate whether additional (possibly more phenomenological) criteria can restrict the number of possible physical metrics further, as one would expect only a single metric to be physical. 

{\bf Higher derivative.} Is there a reformulation of this model in terms of a theory described by a higher-derivative field equation? For $N=2$ (ZDG) with a single invertible dreibein, it is possible to solve the field equation for one of the dreibeine in terms of the invertible one as an expansion in $1/m^2$ \cite{Bergshoeff:2014eca}. 
The resulting field equation is formulated in terms of a single dreibein, but it contains an infinite number of higher-derivative terms. Furthermore, there is no action principle in terms of a single metric. It is also possible to solve the (restricted) $N=3$ (DDG) field equations in a similar fashion to obtain a single higher-derivative field equation in terms of a single invertible dreibein. The resulting field equation features an infinite sum of terms which have at most six derivatives on the metric constructed from the single invertible dreibein. These equations, however, cannot be derived from an action principle involving a single metric, much like in the ZDG case or as in the recently discussed ``Minimal Massive Gravity'' of \cite{Bergshoeff:20144pca}.

The DDG model discussed here contains the sixth order higher-derivative model of \cite{Afshar:2014ffa} called extended NMG as a scaling limit. Both theories have the same linearized spectrum and are free of Boulware-Deser ghosts, although the parameter space of DDG is larger. The details of this scaling limit is presented in the appendix \ref{sec:ENMGlimit}. The physical quantities such as the AdS central charge in the extended NMG model can be found as a limit from the corresponding DDG quantities. 
Although this scaling limit preserves the number of degrees of freedom, it does not preserve unitarity; after taking the scaling limit, it is impossible to satisfy positivity of the masses, energies and the central charge at the same time. The sector of DDG containing extended NMG lies outside the unitary sector of DDG.

\subsection*{Acknowledgments}
We thank Maria Irakleidou, Mehmet Ozkan, Jan Rosseel and Friedrich Sch\"oller for discussion. W. M. wishes to thank the Centre for Theoretical Physics of the University of Groningen for providing an inspiring work environment for the past few years.
H. R. A. and W. M. are supported by the Dutch stichting voor Fundamenteel Onderzoek der Materie (FOM).

\appendix

\section{Hamiltonian Analysis of DDG}
\label{app:Hamil}

In this appendix we give the details on the Hamiltonian analysis of the drei-dreibein gravity model. The calculation is performed along the lines of the Hamiltonian analysis of general Chern-Simons--like models involving only Lorentz-vector valued one-forms as presented in \cite{Hohm:2012vh}. For more details on the general model we refer to there and \cite{Bergshoeff:2014bia}. 

The DDG model is described by a Lagrangian:
\begin{equation}\label{Lgeneral}
\cL = \frac12 g_{rs}  a^r \cdot da^s + \frac16 f_{rst} a^r \cdot (a^s \times a^t)\,,
\end{equation}
where $a^{r}$ is a set of six Lorentz-vector valued one-forms labeled by `field space' indices $r, s, t, \cdots$, which describe the fields $(\omega_1, \omega_2, \omega_3, e_1, e_2, e_3)$. After omitting an overall factor $M_P$, the symmetric field space metric $g_{rs}$ has non-zero entries for:
\begin{align}
g_{e_1\omega_1} = -\sigma_1 \,, && g_{e_2\omega_2} = - \sigma_2 \,, && g_{e_3\omega_3} = -\sigma_3 \,.
\end{align}
The non-zero entries for the symmetric field space matrix $f_{rst}$ are
\begin{align}
& f_{e_1 \omega_1 \omega_1} = -\sigma_1 \,, && f_{e_2 \omega_2 \omega_2} = -\sigma_2 \,, && f_{e_3 \omega_3 \omega_3} = - \sigma_3 \,, \nonumber \\
& f_{e_1e_1e_1} = -\alpha_1 m^2\,, &&  f_{e_2e_2e_2} = -\alpha_2 m^2\,, && f_{e_3e_3e_3} = -\alpha_3 m^2\,,   \nonumber \\ 
& f_{e_1e_1e_2} = \beta_{12}m^2\,, && f_{e_1e_2e_2} = \beta_{21}m^2\,, && f_{e_1e_1e_3} = \beta_{13} m^2\,, \\  \nonumber
& f_{e_1e_3e_3} = \beta_{31}m^2\,, && f_{e_2e_2e_3} = \beta_{23}m^2\,, && f_{e_2e_3e_3} = \beta_{32} m^2\,, \\ \nonumber
&  && f_{e_1e_2e_3} = \beta_{123} m^2\,.
\end{align}
The Lagrangian \eqref{Lgeneral} is a first order Lagrangian and after a space-time decomposition of the fields, the corresponding Hamiltonian is solely a function Lagrange multipliers (the time-components of $a^{r\,a}$) and a set of constraints $\phi_r{}^a$ (see \cite{Hohm:2012vh} or \cite{Bergshoeff:2014bia} for explicit expressions of the constraints). Since the time-components do not propagate, only the spatial parts of the fields contribute to the dynamical phase-space.

From the equations of motion of \eqref{Lgeneral} a set of conditions can be derived which must hold on-shell. They are
\begin{equation}\label{Intcon}
f^t{}_{q[r}f_{s]pt}a^{r}\, a^p \cdot a^q = 0\,.
\end{equation}
Where the index of $f^t{}_{qr}$ is raised with the inverse of $g_{rs}$. eqn.~\eqref{Intcon} are six three-form equations from which we can derive the secondary constraints if they are a function of solely an invertible field. Three of these equations give the Cartan identities \eqref{CartanDDG} which we analyse in section \ref{sec:constraints}. Assuming only invertibility of $e_1$ led to a unique choice where two secondary constraints on the spatial components of $a^{r}$ could be derived from the identities \eqref{CartanDDG}. This parameter restriction was to take only $\beta_{12}$ and $\beta_{13}$ non-zero.\footnote{For two or three invertible fields there are three inequivalent choices of two non-zero coupling constants leading to secondary constraints, as was shown in section \ref{sec:constraints}. One of these three possibilities is worked out here. The Hamiltonian analysis for the other two choices is similar to the analysis presented here and yields the same results.} 
The corresponding secondary constraints are given in \eqref{seccon1}. For this choice of parameters, the other three equations in \eqref{Intcon} reduce to:
\begin{equation}
\begin{split}
\beta_{12} e_2\, \omega_{12}\cdot e_1 + \beta_{13} e_3\, \omega_{13} \cdot e_1 = 0 \,, \\
 e_1\, \omega_{12} \cdot e_1 = 0 \,, \qquad 
 e_1\, \omega_{13} \cdot e_1 = 0\,.
 \end{split}
\end{equation}
These equations lead to another two secondary constraints, given in \eqref{seccon2}. To check the consistency of the primary constraints $\phi_r^a$ under time evolutions, we calculate $d\phi (\xi)/dt$, where $\phi(\xi)$ is a smeared operator defined by integrating $\phi_r^a$ against a vector field $\xi_a^r$. This amounts to calculating the matrix of Poisson brackets \cite{Hohm:2012vh}
\begin{equation} \label{gen_poissonbr}
\left\{  \phi(\xi) , \phi(\eta)  \right\}_{\rm P.B.} =  \phi([\xi, \eta]) + \int_{\Sigma} \xi^r_a \eta^s_b \, \cP_{rs}^{ab} \,,  
\end{equation}
with
\begin{align}
& \cP_{rs}^{ab}  = f^t{}_{q[r} f_{s] pt} \eta^{ab} \Delta^{pq}  +  2f^t{}_{r[s} f_{q]pt} (V^{ab})^{pq}\,, \label{Pmat_def} \\
 & V_{ab}^{pq} =  \varepsilon^{ij} a^p_{i\, a} a^q_{j\, b}\,, \text{  and }\;\; \Delta^{pq} = \varepsilon^{ij} a_i^p \cdot a_j^q\,. \nonumber \\
& [\xi ,\eta]^t_c = f_{rs}{}^{t} \epsilon^{ab}{}_{c} \xi^r_a \eta^s_b \nonumber
\end{align}
By virtue of our parameter choice and the secondary constraints, the first term in \eqref{Pmat_def} is identically zero. The remaining term gives a $18 \times 18$ matrix, $\cP_{rs}^{ab}$ whose entries are given by
\begin{align}\label{Pmat}
(\cP_{ab})_{rs} = & \; m^2 \beta_{12}
\left(
\begin{array}{cccccc}
 0 & 0 & 0 & V_{ab}^{e_1e_2} & - V_{ab}^{e_1e_1} & 0 \\
 0 & 0 & 0 & -V_{ab}^{e_1e_2} & V_{ab}^{e_1e_1} & 0 \\
 0 & 0 & 0 & 0 & 0 & 0 \\
  V_{ab}^{e_2e_1} & - V_{ab}^{e_2e_1} & 0 & - (V_{[ab]}^{\omega_1e_2}-V_{[ab]}^{\omega_2e_2}) &
   V_{ab}^{\omega_1e_1} - V_{ab}^{\omega_2e_1} & 0 \\
 - V_{ab}^{e_1e_1} &  V_{ab}^{e_1e_1} & 0 & V_{ab}^{e_1\omega_1}-V_{ab}^{e_1\omega_2} & 0 & 0 \\
  0 & 0 & 0 & 0 & 0 & 0
\end{array}
\right) \\ 
& + m^2 \beta_{13}
\left(
\begin{array}{cccccc}
 0 & 0 & 0 & V_{ab}^{e_1e_3} & 0 &  - V_{ab}^{e_1e_1} \\
  0 & 0 & 0 & 0 & 0 & 0 \\
 0 & 0 & 0 & - V_{ab}^{e_1e_3} & 0 & V_{ab}^{e_1e_1} \\
 V_{ab}^{e_3e_1} & 0 & - V_{ab}^{e_3e_1} & - (V_{[ab]}^{\omega_1e_3}-V_{[ab]}^{\omega_3e_3}) & 0 & V_{ab}^{\omega_1 e_1}-V_{ab}^{\omega_3 e_1} \\
   0 & 0 & 0 & 0 & 0 & 0 \\
- V_{ab}^{e_1e_1} & 0 & V_{ab}^{e_1e_1} &  V_{ab}^{e_1 \omega_1} -  V_{ab}^{e_1 \omega_3} &
0 & 0
\end{array}
\right)\,. \nonumber
\end{align}
We can determine the rank of this matrix at any point in spacetime by an arbitrary parametrization of the fields and plugging it into Mathematica. We find that this matrix has rank 8. To complete the analysis we must add the Poisson brackets of the primary constraints with the secondary ones. We define
\begin{equation}
\psi_1 = \Delta^{e_1e_2}\,, \quad \psi_2 = \Delta^{e_1e_3} \,, \quad \psi_3 = \Delta^{\omega_1e_1}- \Delta^{\omega_2 e_1}\,, \quad \psi_4 = \Delta^{\omega_1e_1} - \Delta^{\omega_3e_1}\,.
\end{equation}
The Poisson brackets of the secondary constraints among themselves vanish on the constraint surface and the brackets with the primary constraints are given by:
\begin{subequations}
\begin{align}
\{\psi_1, \phi[\xi]\} = & \; \varepsilon^{ij}\left(e_{1\,i} \cdot \partial_{j} \xi^{e_2} - e_{2\,i} \cdot \partial_{j} \xi^{e_1} - (\xi^{\omega_1} - \xi^{\omega_2}) \cdot  e_{1\,i} \times e_{2\,j} - \xi^{e_1} \cdot \omega_{1\,i} \times e_{2\,j} \right. \nonumber  \\
& \left. +\xi^{e_2} \cdot \omega_{2\,i} \times e_{1\,j} \right) \,, 
\\
\{\psi_2, \phi[\xi]\} = & \; \varepsilon^{ij}\left(e_{1\,i} \cdot \partial_{j} \xi^{e_3} - e_{3\,i} \cdot \partial_{j} \xi^{e_1} - (\xi^{\omega_1} - \xi^{\omega_3}) \cdot e_{1\,i} \times e_{3\,j} - \xi^{e_1} \cdot \omega_{1\,i}\times  e_{3\,j} \right. \nonumber  \\
& \left. +\xi^{e_3}\cdot \omega_{3\,i} \times e_{1\,j}  \right) \,, 
\\
\{ \psi_3, \phi[\xi] \} = & \; \varepsilon^{ij} \Big( - e_{1\,i} \cdot \partial_j( \xi^{\omega_1} - \xi^{\omega_2})  + (\omega_{1\,i} - \omega_{2\,i}) \cdot \partial_j \xi^{e_1} 
 \nonumber \\ \nonumber
& + \xi^{e_1} \cdot (\omega_{1\,i} - \omega_{2\,i}) \times \omega_{1\,j} + m^2 \left(\sigma_1^{-1} \beta_{12} \xi^{e_1} + \sigma_2^{-1} \alpha_2 \xi^{e_2} \right) \cdot e_{1\,i} \times e_{2\,j}   \\
& - m^2\left( ( \beta_{12} \sigma_2^{-1} + \alpha_1 \sigma_1^{-1}) \xi^{e_1} - \beta_{12} \sigma_1^{-1} \xi^{e_2} - \beta_{13} \sigma_1^{-1} \xi^{e_3} \right) \cdot e_{1\,i} \times e_{1\,j} \nonumber \\
&  + \beta_{13} \sigma_1^{-1} m^2 \xi^{e_1} \cdot e_{1\,i} \times e_{3\,j} - (\xi^{\omega_1} - \xi^{\omega_2}) \cdot e_{1\,i} \times \omega_{2\,j} \Big)\,,  \\
\{ \psi_4, \phi[\xi] \} = & \; \varepsilon^{ij} \Big( - e_{1\,i} \cdot \partial_j( \xi^{\omega_1} - \xi^{\omega_3})  + (\omega_{1\,i} - \omega_{3\,i}) \cdot \partial_j \xi^{e_1} 
 \nonumber \\ \nonumber
& + \xi^{e_1} \cdot (\omega_{1\,i} - \omega_{3\,i}) \times \omega_{1\,j} + m^2 \left(\beta_{13} \sigma_1^{-1} \xi^{e_1} +  \alpha_3 \sigma_3^{-1} \xi^{e_3} \right) \cdot e_{1\,i} \times e_{3\,j}   \\
&  - m^2\left( ( \beta_{13} \sigma_3^{-1} + \alpha_1 \sigma_1^{-1}) \xi^{e_1} - \beta_{12} \sigma_1^{-1}  \xi^{e_2} - \beta_{13} \sigma_1^{-1} \xi^{e_3} \right) \cdot e_{1\,i} \times e_{1\,j} \nonumber\\
& + \beta_{12} \sigma_1^{-1} m^2 \xi^{e_1} \cdot e_{1\,i} \times e_{2\,j} - (\xi^{\omega_1} - \xi^{\omega_3}) \cdot e_{1\,i} \times \omega_{3\,j} \Big)\,.
\end{align}
\end{subequations}
For general values of the coupling constants, adding these brackets to the total matrix of Poisson brackets will increase the rank of that matrix by 8, making a $22 \times 22$ matrix of rank 16. This implies that there are $22-16= 6 $ first class constraints, consistent with the number of gauge symmetries in the theory. The remaining 16 constraints are second class. This leads to the degree of freedom count as
\begin{eqnarray}
\# \text{ d.o.f.} = \frac12 \left( 6 \times 3 \times 2 - 16 - 2 \times 6\right) = 4\,.
\end{eqnarray}
This result is consistent with the linear analysis of the DDG model, which propagates two massive spin-2 particles, each with two helicity states.

\section{Scaling limit to extended NMG}\label{sec:ENMGlimit}
The extended NMG theory obtained in \cite{Afshar:2014ffa} has the same linearized spectrum as the ghost-free drei-dreibein gravity, although the parameter space of the latter is larger. In fact, there exists a scaling limit, or a flow, from DDG to the extended NMG theory. Consider the DDG Lagrangian \eqref{DDG} with the same ghost-free parameter choice as was discussed in section \ref{LinDDG};
\begin{equation}\label{LDDGlim}
\begin{split}
L_{\text{\tiny{DDG}}}= - M_P  \bigg\{ & \sigma_1 e_{1} \cdot R_1 + \sigma_2  e_{2} \cdot R_2 + \sigma_3 e_{3} \cdot R_3 \\ 
& + \frac{m^2}{6}  \big( \alpha_1 e_1 \cdot e_1 \times e_1 +\alpha_2 e_2 \cdot e_2 \times e_2 + \alpha_3e_3 \cdot e_3 \times e_3\big) \\
&  - \frac{m^2}{2} \big(\beta_{12} e_1 \cdot e_1\times e_2 +\beta_{23} e_2 \cdot e_2 \times e_3 \big) \bigg\}\,.
\end{split}
\end{equation}
We introduce the following parametrization as an expansion in $\lambda$ for the three dreibeine and the spin connections,
\begin{eqnarray}\label{limit}
  \begin{split}
    e_1 &= e\,,\\
    e_2 &= e +\tfrac{\lambda}{m^2} f_1\,,\\
    e_3 &= a_{31}\,e+a_{32}\tfrac{\lambda}{m^2}  f_1 +\tfrac{\lambda^2}{m^4}  f_2\,,
  \end{split}
\qquad\qquad
  \begin{split}
    \omega_1 &= \omega\,,\\
    \omega_2 &= \omega +\tfrac{\lambda}{m^2}  h_1\,,\\
    \omega_3 &= \omega +\tfrac{a_{32}}{a_{31}}\tfrac{\lambda}{m^2}  h_1+ \tfrac{1}{a_{31}}\tfrac{\lambda^2}{m^4}  h_2\,.
  \end{split}
\end{eqnarray}
For shorthand notation we may rename the parameters as $a_{31}\equiv A$ and $a_{32}\equiv B$. 
We take the Planck mass and the $\sigma_I$ parameters as
\begin{equation}
M_P= \frac{M}{\lambda^2}\,, \qquad \sigma_1 = B - A + \lambda^2 \sigma \,,\qquad \sigma_2 = -B\,,\qquad \sigma_3 = 1\,,
\end{equation}
where $\sigma=\pm1$ is a new sign parameter. The cosmological parameters and the two coupling constants are expanded as
\begin{align}
& \alpha_1 = -2 + \frac{2(A-B)}{\lambda}\,, \qquad
\alpha_2 = - 1 + \left(A-B+\frac{2A^2}{B-A}\right)\frac{1}{\lambda} \,,\qquad 
\alpha_3 = \frac{1}{A(B-A)\lambda}\,, \nn \\
&\beta_{12} = \frac{\Lambda_0\lambda^2}{3 m^2 }- 1 +\frac{A-B}{\lambda} \,, \qquad 
\beta_{23} = \frac{A}{(B-A)\lambda}\,.
\end{align}
After plugging this into the Lagrangian \eqref{LDDGlim} and taking the limit  $\lambda\rightarrow0$ we arrive at the extended new massive gravity Lagrangian 3-form \cite{Afshar:2014ffa},
\begin{equation}
\begin{split}
L _{\text{\tiny{ENMG}}} & = M \bigg\{ -\sigma e \cdot R + \frac{\Lambda_0}{6} e \cdot e \times e + \frac{1}{2m^2} e \cdot f_1 \times f_1  -\frac{1}{m^4}\Big[e  \cdot \cD h_2    \\
 & + \frac{a}{6} f_1 \cdot f_1 \times f_1 + f_{2} \cdot \left( R + e \times f_{1} \right) + b\,  h_{1} \cdot \left(\cD f_1 + \tfrac{1}{2} e \times h_{1} \right)\Big] \bigg\} \,.
\end{split}
\end{equation}
where, $a=\frac{B^2-A^2}{A}$ and $b=\frac{B^2-AB}{A}$. After integrating out the auxiliary fields $f_{1,2}$ and $h_{1,2}$ and going to a metric formalism, this Lagrangian describes a sixth-order derivative theory of gravity with the Lagrangian density
\begin{equation} \label{NMG2}
 \mathcal{L}_{\text{\tiny{ENMG}}} = \sigma R-2\Lambda_0 + \frac{1}{m^2}\left( R_{\mu\nu}R^{\mu\nu} - \frac38 R^2 \right) + \frac{1}{m^4}\left( 2 a \det(S) - b \,C^{\mu\nu}C_{\mu\nu}\right)\,.
\end{equation}
Here $\det(S)$ is the determinant of the Schouten tensor $S_{\mu\nu} = R_{\mu\nu} + \frac14 g_{\mu\nu} R$ and $C_{\mu\nu} = \det(e)^{-1}\epsilon_{\mu}{}^{\alpha\beta} \nabla_{\alpha} S_{\beta\nu}$ is the Cotton tensor.

In order to take the limit in the DDG central charge \eqref{DDGcc} to find the AdS  central charge of  extended NMG, we need to know how $\gamma_2$ and $\gamma_3$ in \eqref{fluctuations} scale with $\lambda$. We can deduce this from the parameter relations \eqref{gammas} which in this case read
\begin{align}\label{gammas2}
\sigma_1 \frac{\Lambda}{m^2}  & =   2\beta_{12}\gamma_2 - \alpha_1  \,, \nn \\
 \sigma_2 \frac{\Lambda}{m^2}  &=   \beta_{12} + 2\beta_{23}\gamma_2\gamma_3 - \gamma_2^2 \alpha_2\,, \\
 \sigma_3 \frac{\Lambda}{m^2} &=  \beta_{23}\gamma_2^2 - \alpha_3\gamma_3^2 \,. \nn
\end{align}
If we expand the cosmological constant as $\Lambda = - 1/\ell^2 + \Lambda^{(1)} \lambda + \cO (\lambda^2)$, then it is possible to solve these equations order by order in $\lambda$. The result is
\begin{equation}
\begin{split}
\gamma_2 & =1 + \frac{1}{2\ell^2 m^2 } \lambda + \left( \frac{1}{2(A - B)m^2\ell^2} - \frac{\Lambda^{(1)}}{2m^2}\right) \lambda^2  +\mathcal{O}(\lambda^3)\,,\\
\gamma_3 & = A + \frac{B}{ 2\ell^2 m^2} \lambda - \left( \frac{a}{8\ell^4m^4} + \frac{B}{2b\ell^2m^2} + \frac{B\Lambda^{(1)}}{2}\right)\lambda^2 +\mathcal{O}(\lambda^3)\,.
\end{split}
\end{equation}
Taking the $\lambda \to 0$ limit in the DDG central charge \eqref{DDGcc}, we find that
\begin{equation}
c_{L/R} = \frac{3 \ell}{2G} \left( \sigma +\frac{1}{2\ell^2m^2}-\frac{a}{8\ell^4m^4} \right) \,,
\end{equation}
which agrees with the result obtained in \cite{Afshar:2014ffa}.

%

\providecommand{\href}[2]{#2}\begingroup\raggedright\endgroup

\end{document}